\documentclass[twoside,slac_one]{revtex4}
\usepackage{graphicx}
\usepackage{fancyhdr}
\usepackage{amsmath} 
\usepackage{bm}
\usepackage{amsxtra}
\usepackage{amssymb}
\usepackage{amsthm}
\usepackage{latexsym}
\usepackage{lscape}

\begin{document}

\title{Sound Propagation on Top of the Fireball}

%

\author{P. Staig\footnote{Speaker}}
\author{E. Shuryak}
\affiliation{Department of Physics and Astronomy, Stony Brook University, Stony Brook, NY, USA}

\begin{abstract}
We study the effect that initial state fluctuations have on final particle correlations in heavy ion collisions. More precisely, we focus on the propagation of initial perturbations on top of the expanding fireball using the conformal solution derived by Gubser and Yarom for central collisions. For small perturbations, the hydrodynamic equations are solved by separation of variables and the solutions for different modes are added up to construct initial point-like perturbations, that are then allowed to evolve until freeze-out. The Cooper-Frye prescription is used to determine the final particle distribution. We present the two-particle correlation functions and their Fourier spectra obtained for different viscosities and different initial widths of the perturbation. We find that viscosity kills the higher harmonics, but that the Fourier spectra presents maxima and minima, similar to what is seen in the study of cosmic microwave background radiation. 
\end{abstract}

\maketitle

\thispagestyle{fancy}


\section{Introduction}
The main idea behind the present work  was to study the evolution of an initial perturbation on top of the background produced in heavy ion collisions, and to calculate measurable quantities such as the two particle correlations and the Fourier coefficients of its expansion. However, before going into any of the details of our study we would like to start by talking about the background of the topic to be discussed and by giving some motivation for our calculations. \\

When two relativistic heavy ions collide at experiments such as the Relativistic Heavy Ion Collider (RHIC) and the Large Hadron Collider (LHC) a new form of matter, the quark gluon plasma (QGP), is created. There is evidence \cite{Adams:2005dq,Adcox:2004mh,Arsene:2004fa,Back:2004je,Aamodt:2010pa} for the  existence of a large elliptic flow -characterized by the Fourier coefficient $v_2$- which is in agreement with hydrodynamical simulations, and thus supports the idea that the QGP behaves like a fluid and  can be studied using hydrodynamics. For a long time hydrodynamic studies used smooth initial distributions from which the $v_2$ coefficient could be calculated but which due to symmetry considerations implied that all the odd coefficients in the Fourier expansion of the particle distributions were zero. The work by Alver and Roland \cite{Alver:2010gr}, where initial fluctuations due to the random positions of the nucleons in the nuclei were included, showed that the $v_3$ coefficient was of crucial importance for understanding the ridge and the double hump structure in two particle correlations.  The $v_3$ and other large harmonics have now been measured experimentally \cite{Steinberg:2011dj} for a wide range of centralities and this together with the fact that the $v_2$ coefficient is present at even the most central collisions strengthens the importance of the understanding of the initial state and its fluctuations.\\

In a previous work \cite{Staig:2010pn} we used the Glauber model to compute the initial deformations $\epsilon_n$ and angles $\psi_n$ up to $n=6$, and found that there existed correlations between all odd harmonics, $n=1,3,5$. We understood these correlations to be evidence for the presence of hot or cold spots; initial perturbations to the background that will propagate during the dynamical evolution of the matter created in the collision. In this talk we will discuss the idea of sound  propagating from these initial hot spots.  The idea was introduced in \cite{Shuryak:2009cy} and followed up in \cite{Staig:2010pn} and \cite{Staig:2011wj}, and consists mainly in sound circles that propagate from the initial perturbation and expand until freeze-out reaching the ``sound horizon'' radius. The excess of matter due to these circles would give rise to a particle distribution with two peaks or horns which has also been observed in the calculations done in \cite{Andrade:2009em}.\\

This talk is mostly based on work done in \cite{Staig:2011wj}, where we studied the evolution of an initial perturbation on top of the flow developed in \cite{Gubser:2010ze,Gubser:2010ui} by S. Gubser and A. Yarom. To simplify our calculations and to gain understanding of the way in which perturbations propagate, and their effects on measurable quantities such as the two particle correlations, we study the case with only one Gaussian perturbation on top of the background temperature. Using the tools from \cite{Gubser:2010ui} the initial state and its hydrodynamical evolution can be described analytically.  It is important to notice that in order to do this we we will need to restrict ourselves to consider only central collisions with conformal matter satisfying $\epsilon=3p$ and thus with a fixed speed of sound during the whole evolution.  We will also need to limit ourselves to small perturbations since we will be using linearised solutions to the hydrodynamic equations.\\  

The plan that we will follow in this talk is to start by placing a Gaussian shaped initial perturbation on top of the background and allow it to evolve until freeze-out, where using the Cooper-Frye formula we obtain the particle distribution and from there the two particle correlation  and the coefficients of its Fourier expansion can be calculated and then  compared to experimental data.

\section{Perturbations on Top of the Fireball}
In this section we present the evolution of an initial perturbation on top of the background flow  but, before going into this, it is necessary to understand the background, so we give a summary of Gubser's flow. 
\subsection{Summary of Gubser's Flow}
Let us briefly discuss Gubser's flow which was developed in \cite{Gubser:2010ze} as a generalization to Bjorken flow \cite{Bjorken:1982qr} and extended in \cite{Gubser:2010ui}. The assumptions made by Bjorken are as follows: boost invariance along the beamline, rotation invariance in the azimuthal angle $\phi$ around the beamline  and translation invariance in the transverse plane. The first two of these premises are maintained by Gubser, but the third one is modified by a special conformal transformation in order to take into account the finite size and the transverse expansion of the fireball. The solution can be found for conformal matter with equation of state $ \epsilon=3p \sim T^4 $, which implies a  speed of sound  $c_s=1/\sqrt{3}$.  In coordinates $(\tau,\,r,\,\phi,\, \eta)$ the four-velocity is given by
\begin{eqnarray}
u_{\mu} &  = &
\left(-\cosh{\kappa(\tau,r)},\sinh{\kappa(\tau,r)},0,0\right),
\end{eqnarray}
with
\begin{eqnarray}
v_\perp & = & \tanh{\kappa(\tau,r)}  =  \left(\frac{2q^2\tau
r}{1+q^2\tau^2 + q^2r^2}\right),
\end{eqnarray}
while the energy density corresponds to
\begin{eqnarray}
\epsilon & = & \frac{\hat{\epsilon}_0 (2
q)^{8/3}}{\tau^{4/3}\left(1+2q^2(\tau^2 +
r^2)+q^4(\tau^2-r^2)^2\right)^{4/3}}.
\end{eqnarray}
Here $\hat{\epsilon}_0$ is just a normalization constant and $q$, which has units of inverse length, is the parameter  that accounts for the finite size of the nucleus and for the radial flow. When q goes to zero the transverse velocity vanishes and Bjorken's solution is recovered. \\

The solution that we have just described can be re-derived if one uses  transformation from the $\tau,r$ coordinates to a new set $\rho,\theta$ given by:
\begin{eqnarray}
\sinh{\rho} & = & -\frac{1-q^2\tau^2+q^2r^2}{2q\tau}\label{rho_coord}\\
\tan{\theta} & = &
\frac{2qr}{1+q^2\tau^2-q^2r^2},\label{theta_coord}
\end{eqnarray}
and re-scales the metric as follows
\begin{eqnarray}
ds^2 & = & \tau^2 d\hat{s}^2.
\end{eqnarray}
This was done in  \cite{Gubser:2010ui} by  Gubser and Yarom, and has the advantage that the fluid is now at rest (the only non-zero component of the four-velocity is $u_{\rho}=-1$), so expressed in this new way the solution looks much simpler, and as we will see in the following section it is possible to analytically study perturbations on top of the background that it describes. The rescaled metric in the new coordinates reads:
\begin{eqnarray}
d\hat{s}^2 & = &-d\rho^2 + \cosh^2{\rho}\left(d\theta^2 +
\sin^2{\theta}d\phi^2\right)+d\eta^2,
\end{eqnarray}
where, the new time coordinate is  $\rho$  and the new radial coordinate is $\theta$. For the the case with shear viscosity $\eta$ the temperature in the rescaled frame is 
\begin{eqnarray}
\hat{T} & = &\frac{\hat{T}_0}{(\cosh{\rho})^{2/3}} +\frac{H_0
\sinh^3{\rho}}{9 (\cosh{\rho})^{2/3}} \,
_2F_1\left(\frac{3}{2},\frac{7}{6};\frac{5}{2},-\sinh^2{\rho}\right)\nonumber\\
\label{back_T}
\end{eqnarray}
where $\hat{T}_0$ is a normalization constant, $H_0=\eta /T^3$  and $_2F_1$ is
the hypergeometric function. 

\subsection{Perturbations of the Flow}
In real heavy ion collisions the fireball is not smooth  but it presents fluctuations due to the random positions of the nucleons in the nuclei.  In order to take this into account, following the calculations done in \cite{Gubser:2010ui}, we studied perturbations on top of Gubser's flow. Since we are interested in trying to understand the final two-particle correlations which present long range correlations in rapidity, we study only perturbations that are rapidity independent, this is perturbations in the transverse plane. The temperature and the four-velocity in the rescaled metric can be written as the background plus the perturbation in the form:
\begin{eqnarray}
\hat{T} & = &  \hat{T}_b(1+\delta)\label{Tpertb}\\
\hat{u}_{\mu} & = & \hat{u}_{b \,\mu} + \hat{u}_{1\mu}\label{upert}
\end{eqnarray}
where the perturbations correspond to
\begin{eqnarray}
\delta & = & \delta(\rho,\theta,\phi)\\
\hat{u}_{1\mu} & = & (0,u_{\theta}(\rho,\theta,\phi),u_{\phi}(\rho,\theta,\phi),0)
\end{eqnarray}
By assuming that the perturbations are small,  the hydrodynamic equations $\nabla^{\mu}T_{\mu\nu}$ can be treated keeping only linear terms. By doing this one obtains a system of coupled first order differential equations, which in the ideal case may be decoupled to generate the following  second order differential equation for the temperature perturbation   
\begin{eqnarray}
& &\frac{\partial^2 \delta}{\partial \rho^2} -
\frac{1}{3\cosh^2{\rho}} \left( \frac{\partial^2 \delta}{\partial
\theta^2}  +\frac{1}{\tan{\theta}}\frac{\partial \delta}{\partial
\theta}+ \frac{1}{\sin^2{\theta}}\frac{\partial^2 \delta}{\partial
\phi^2}
\right)  
+\frac{4}{3}\tanh{\rho}\frac{\partial \delta}{\partial \rho}=0.
\label{T_pert_eqn}
\end{eqnarray} 
Using separation of variables the solution is found to be
\begin{eqnarray}
\delta(\rho,\theta,\phi)= \delta_l(\rho)Y_{lm}(\theta,\phi)\label{del}
\end{eqnarray}
where the $\delta_l(\rho)$ functions are given by combination of associated Legendre polynomials times $\cosh{(\rho)}^{-2/3}$
and the $Y_{lm}$ are spherical harmonics.
\begin{figure}[ht]
\includegraphics[width=0.32\textwidth]{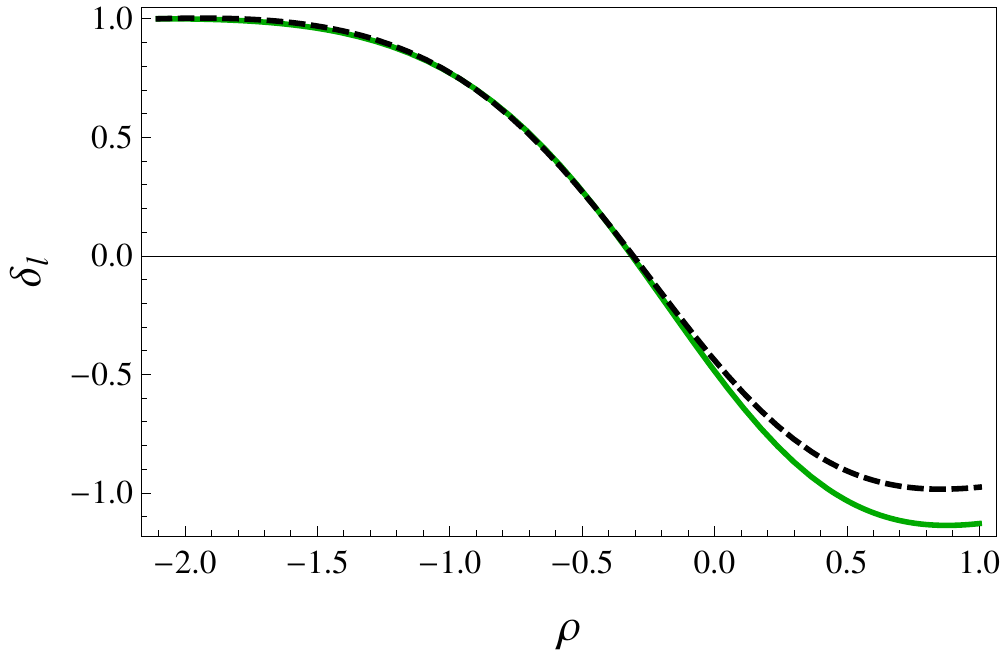}
\includegraphics[width=0.32\textwidth]{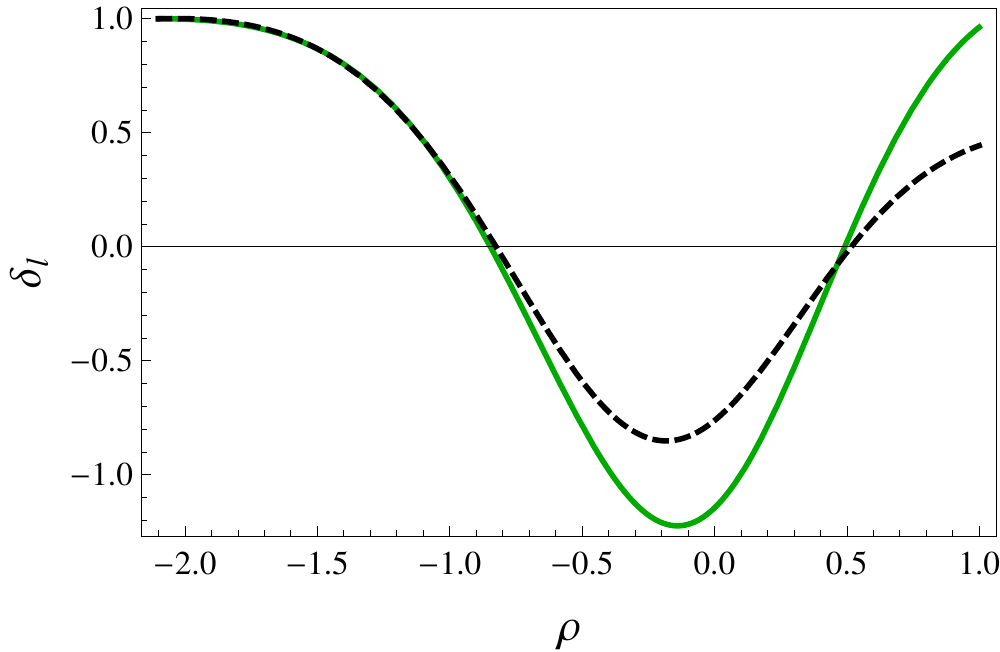}
\includegraphics[width=0.32\textwidth]{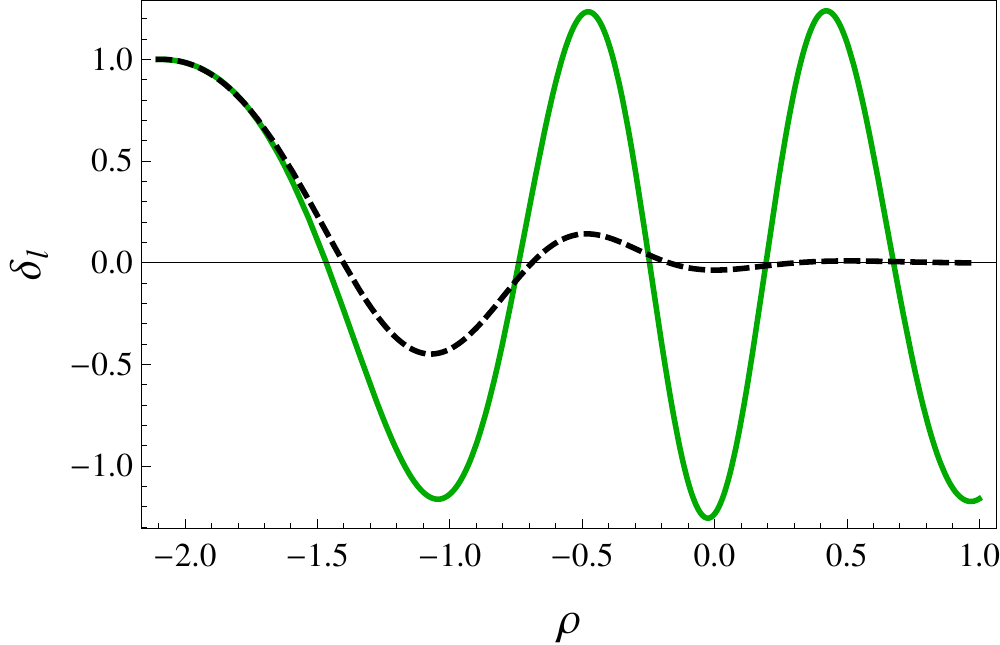}
\includegraphics[width=0.32\textwidth]{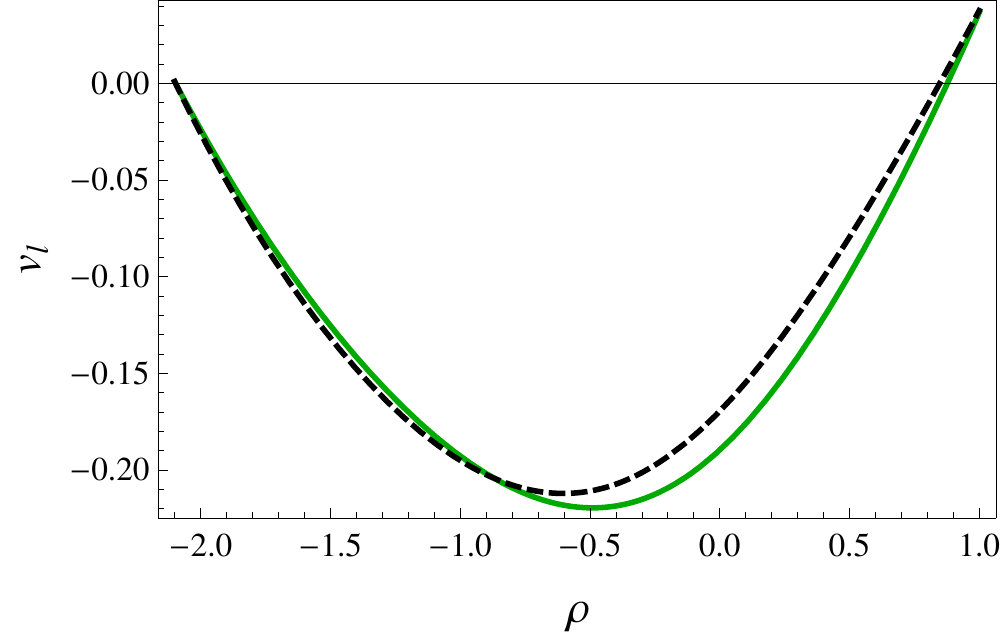}
\includegraphics[width=0.32\textwidth]{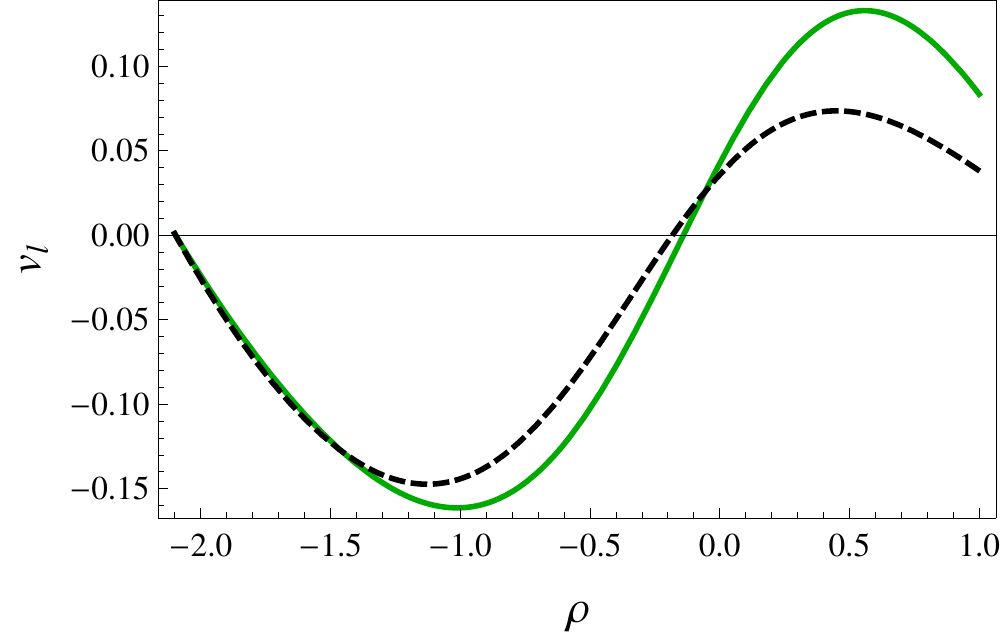}
\includegraphics[width=0.32\textwidth]{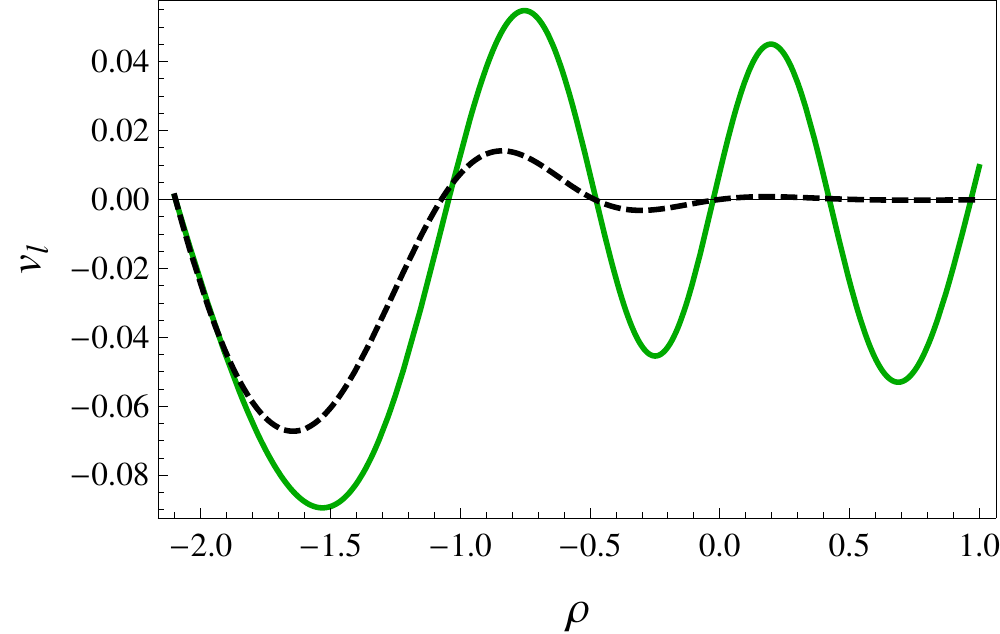}
\caption{Comparison between the magnitude of the $\delta_l$ (top) and of the $v_l$ (bottom) for the ideal case (solid green lines) and the viscous case with $\eta/s=0.16$ (dashed black lines), for l=2,4,12 from left to right.\label{compare}}
\end{figure}
The perturbations to the velocity  are as follows
\begin{eqnarray}
\hat{u}_{\theta} & = & v_l(\rho)\partial_{\theta}Y_{lm}(\theta,\phi) \label{uth}\\
\hat{u}_{\phi} & = & v_l(\rho)\partial_{\phi}Y_{lm}(\theta,\phi)\label{uph}
\end{eqnarray}
with the $\rho$-dependent part given by
\begin{eqnarray}
v_l(\rho) & = & \frac{3\cosh^2{\rho}}{l(l+1)}\frac{d
\delta_{l}}{d\rho}.
\end{eqnarray}
In the viscous case the equations are more complicated, but separation of variables can still be used.  The $\theta, \, \phi$ part of the solution does not change, but the  $\rho$-dependent part does and it needs to be solved numerically in this case. The comparison between the ideal and the viscous case is presented in Fig.\ref{compare} for three different values of l.  These plots show the oscillatory behaviour  of the solutions, which is more evident for larger l; what begins as only a perturbation in temperature then becomes a perturbation in velocity and then keeps oscillating between them.  The effect of viscosity can be seen for larger times, and it consists in  a damping of the amplitude of the oscillation. This damping is greater for larger l, so in the cases where viscosity is present the larger harmonics are suppressed, as we expected from the discussion in section II A of \cite{Staig:2010pn}.\\

\begin{figure}[ht]
\centering
\includegraphics[width=0.32\textwidth]{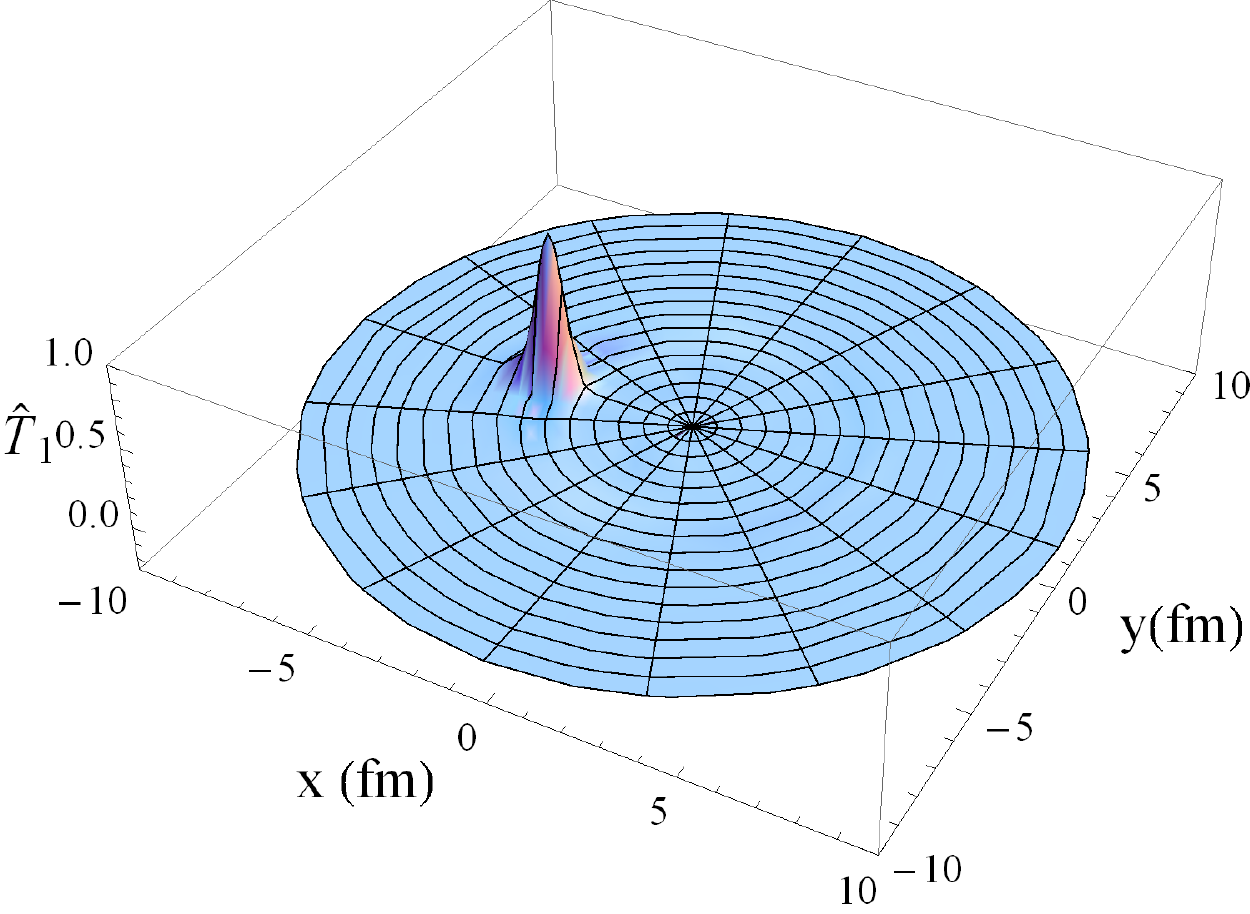}
\includegraphics[width=0.32\textwidth]{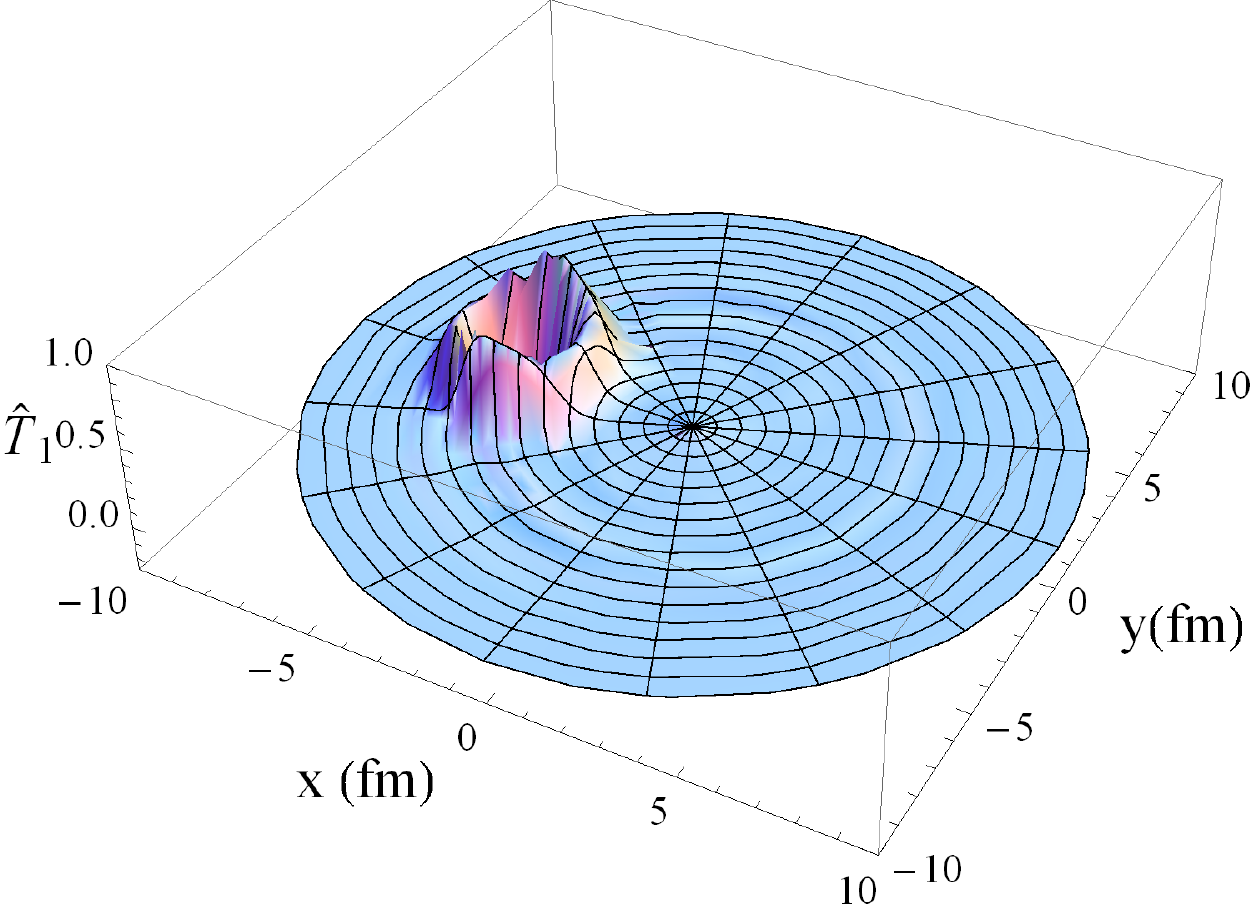}
\includegraphics[width=0.32\textwidth]{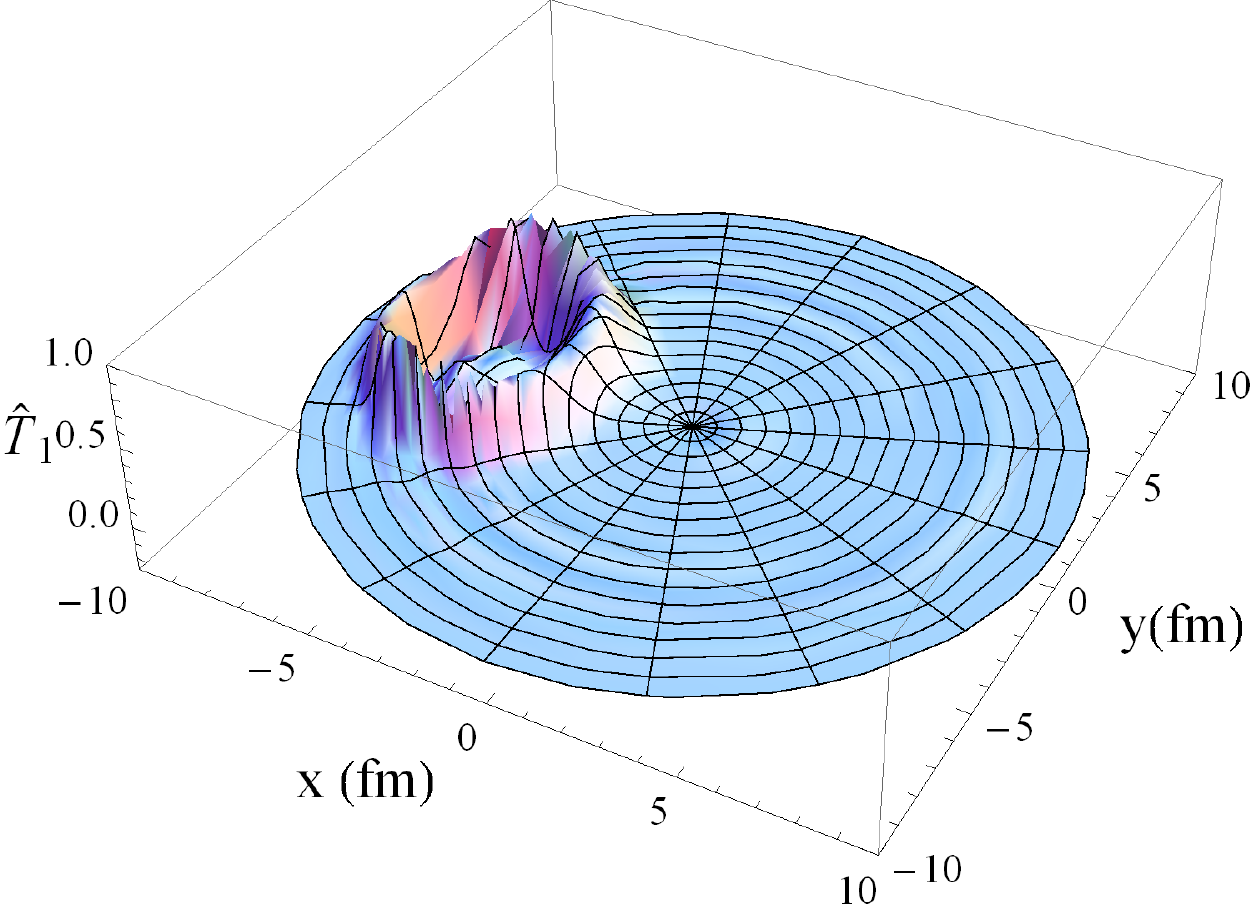}
\caption{Evolution of the temperature perturbation $\hat{T}_1(\tau,r,\phi)$ in the rescaled frame but in the regular coordinates for $\tau=1,4,6\, fm/c$ from left to right.  (Taken from \cite{Staig:2011wj})\label{evolution}} 
\end{figure}

Looking back to Eq.\ref{del}, we see that at any given time $\rho$ the temperature perturbation is given in terms of spherical harmonics of the radial coordinate $\theta$ and the azimuthal coordinate $\phi$, so we can use the completeness of these functions in order to write any initial condition as a sum of them by calculating the appropriate coefficients. In particular, at an initial time $\rho=\rho_0$ we used a Gaussian perturbation in temperature  with width $s$ localized at $\theta=\theta_0$ and $\phi=\phi_0$ given by
\begin{eqnarray}
\hat{T}_1(\rho_0,\theta,\phi) \propto e^{-\frac{\theta^2 +
\theta_0^2 - 2 \theta \theta_0 \cos{(\phi-\phi_0)}}{2
s^2}},\label{IC1}
\end{eqnarray}
and assumed zero initial perturbation to the velocity. By doing this we were able to write the temperature perturbation (in the rescaled frame) at any given time $\rho$ as
\begin{eqnarray}
\hat{T}_1(\rho,\theta,\phi) & = & \sum_l \sum_{m=-l}^{m=l}
 c_{lm} R_l(\rho)Y_{lm}(\theta,\phi)\label{Tpert}
\end{eqnarray}
where $R_l(\rho)=\hat{T}_b(\rho) \delta_l(\rho)$ and the $c_{lm}$ coefficients are calculated by using the orthogonality of the spherical harmonics and the initial conditions.  In this way we may follow the evolution of the perturbation as it is shown in the three plots from Fig.\ref{evolution}.
What we find is that all the harmonics (we used up to $l=30$) add up coherently to form the Gaussian initial perturbation, which then starts to propagate as an expanding circle on top of the background.
\section{The Sound Circle and Final Particle Distributions}
In this section we present the single particle distribution, the two particle correlation and the power spectrum calculated from the perturbation that evolved on top of the background  at freeze-out. 
\subsection{Single Particle Distribution}
In order to calculate the particle distribution we used the Cooper-Frye formula \cite{Cooper:1974mv}
\begin{eqnarray}
E\frac{dN}{d^3p}=-\int d\Sigma_{\mu}p^{\mu}f\left(\frac{p^{\nu}u_{\nu}}{T}\right)
\end{eqnarray}
where the function $f$ corresponds to the thermal distribution, $\Sigma$ is the freeze-out surface and the overall minus is there because we work in the mostly plus metric. In this work we used an isothermal freeze-out, this is: the freeze-out surface was calculated from  $T(\tau,x,y,\eta)=T_{f}$ which was solved to get $\Sigma^{\mu}=(\tau_{f}(x,y),x,y,\eta)$. The shape of the freeze-out surface, and thus the freeze-out time, depends then on the temperature distribution: where the temperature was higher, the freeze-out time will be larger because it will take more time to reach the  temperature of the freeze-out. In this way the effects of the temperature perturbation are preserved, the expanding circle which had a higher temperature than the background will have a larger freeze-out time and it will be visible in the freeze-out surface. The intersection between the edge of the fireball and the circle  give rise to the particle distribution.\\
\begin{figure}[ht]
\centering
\includegraphics[width=0.32\textwidth]{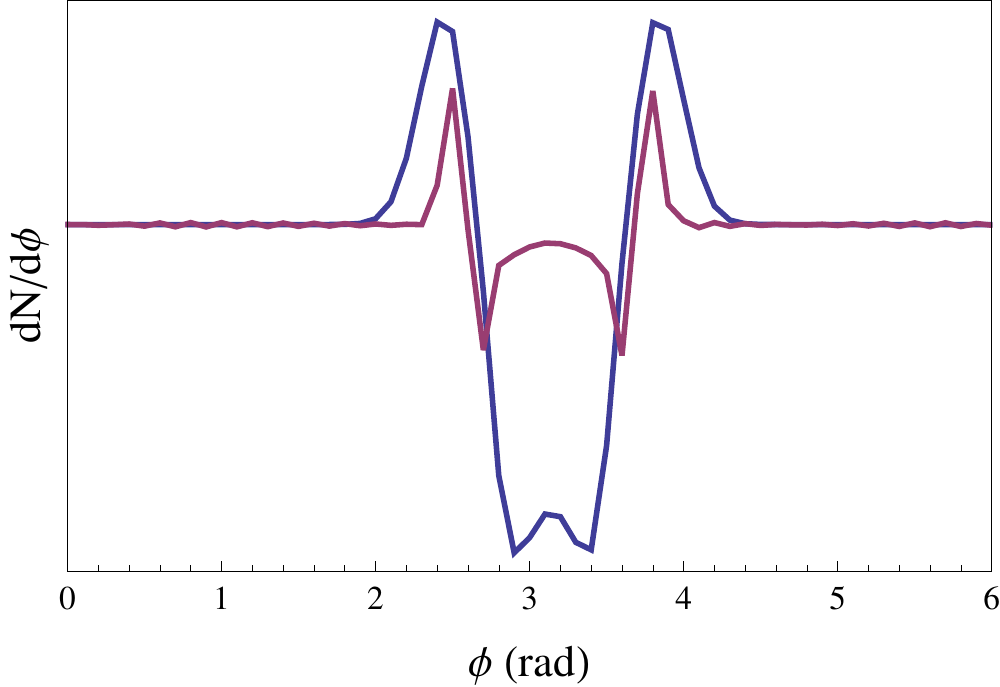}
\includegraphics[width=0.32\textwidth]{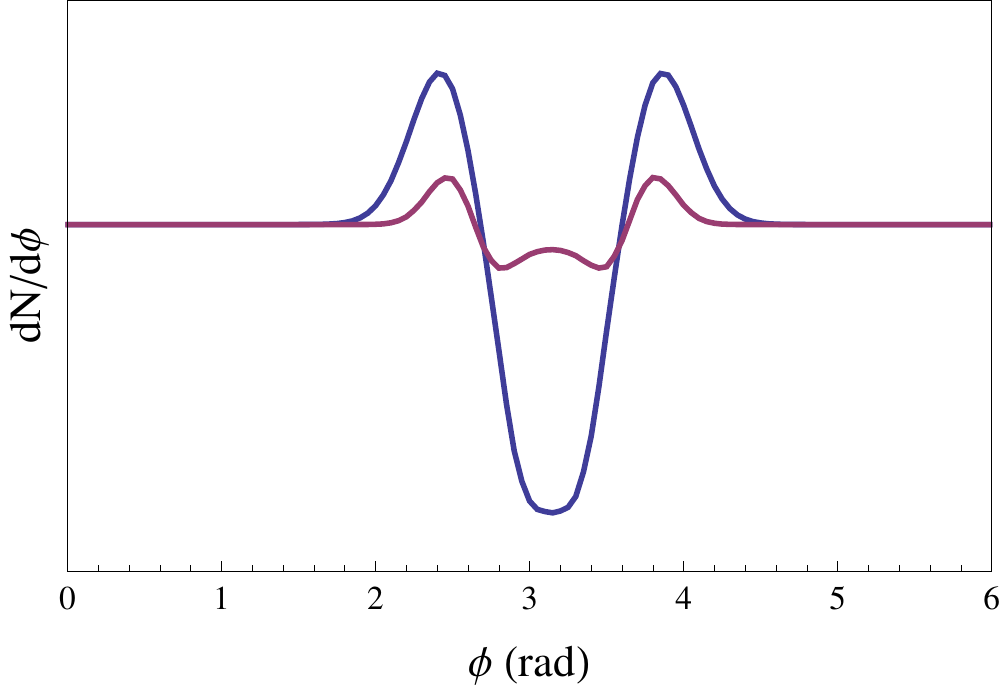}
\includegraphics[width=0.32\textwidth]{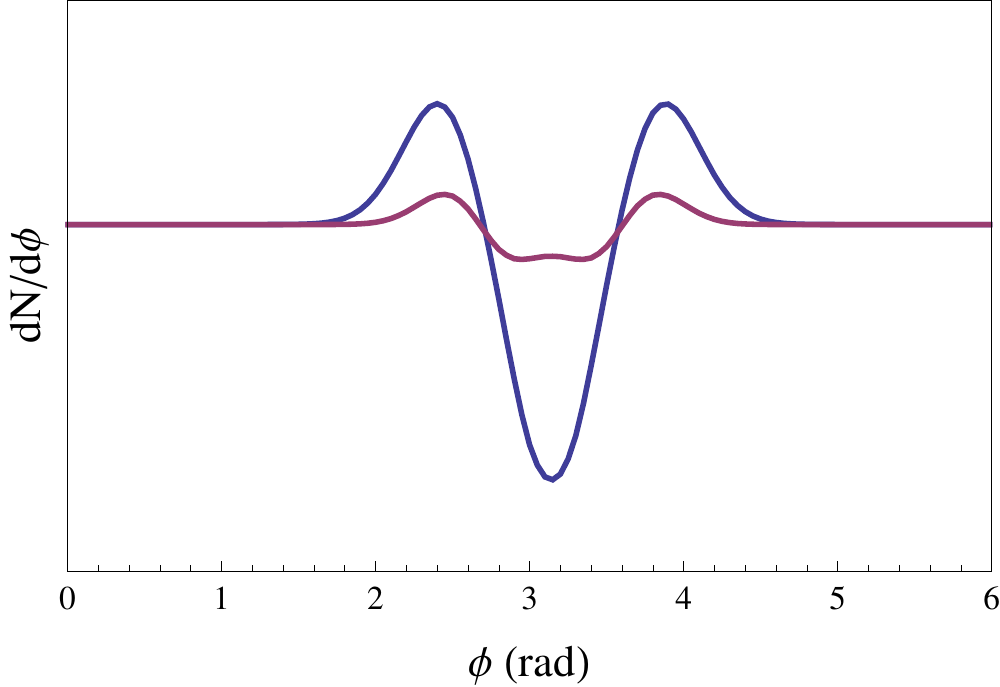}
\caption{Single particle distribution (arbitrary units) for initial Gaussian perturbations with widths 0.4 fm (magenta) and 1 fm (blue), for three different viscosities $\eta/s=0,0.08,0.16$ from left to right.\label{pdist}} 
\end{figure}

The particle distributions as a function of the azimuthal angle $\phi$ obtained for different initial widths of the perturbation and different viscosities are shown in Fig.~\ref{pdist}.  It is important to notice here that we have taken a qualitative approach; we are interested in the shapes of the distributions and in the effects of viscosity and initial width of the perturbation, so we used arbitrary units, which is why we have suppressed the labelling of the y axis. The main features of the distribution are two horns with a dip in the middle which are centred around the initial position of the perturbation. In the ideal case there appears a bump in the middle of the dip, but it becomes less important when viscosity is included. Other  effects of viscosity are to damp and broaden the peaks and smooth out the curves.  The width of the initial Gaussian perturbation also  affects the final particle distribution; for a greater initial width the main characteristics of the curves are more pronounced: the maxima are larger and the minimum in the middle is more pronounced, also the width of the peaks is larger.  This is because we kept the initial height of the perturbation and changed the width, so the total energy of the perturbation is larger, making the effects on the final distribution larger as well.
\subsection{Two Particle Correlations}
\begin{figure}[ht]
\centering
\includegraphics[width=0.4\textwidth]{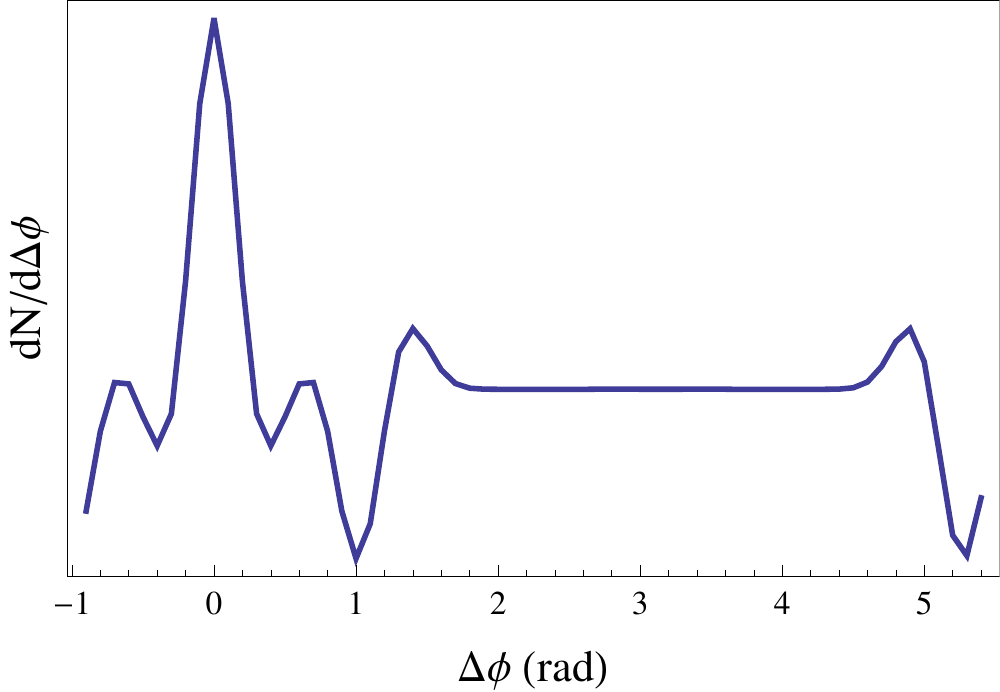}
\includegraphics[width=0.4\textwidth]{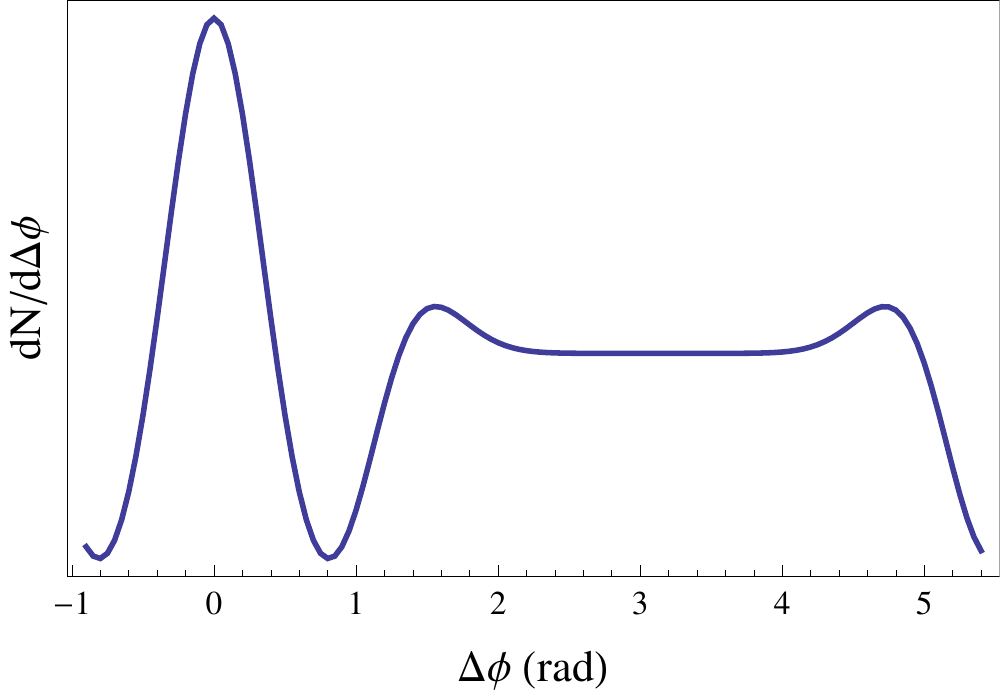}
\includegraphics[width=0.4\textwidth]{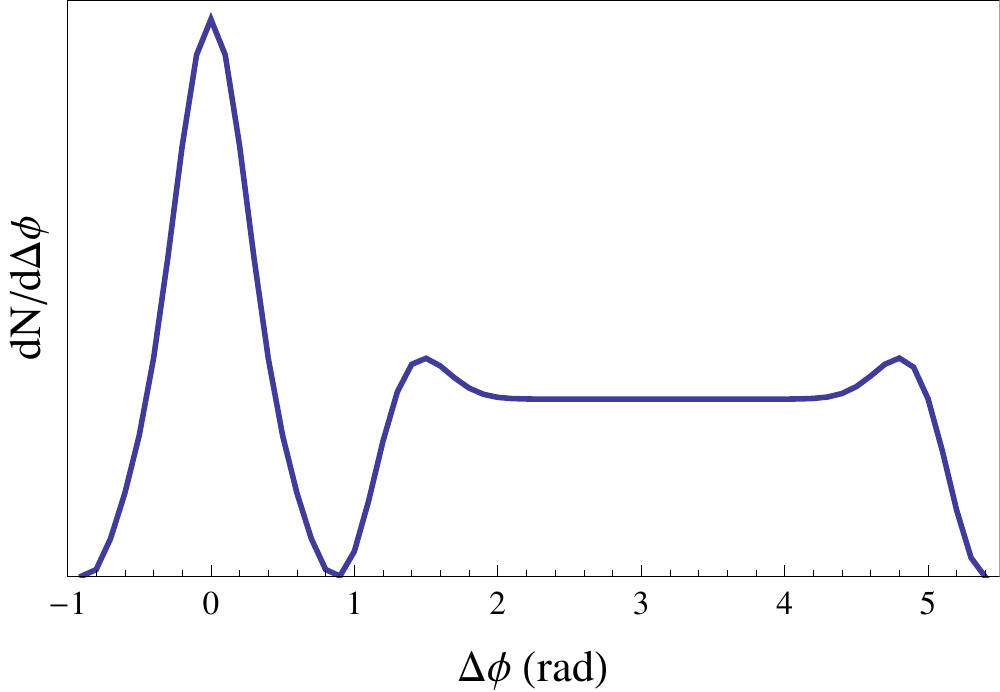}
\includegraphics[width=0.4\textwidth]{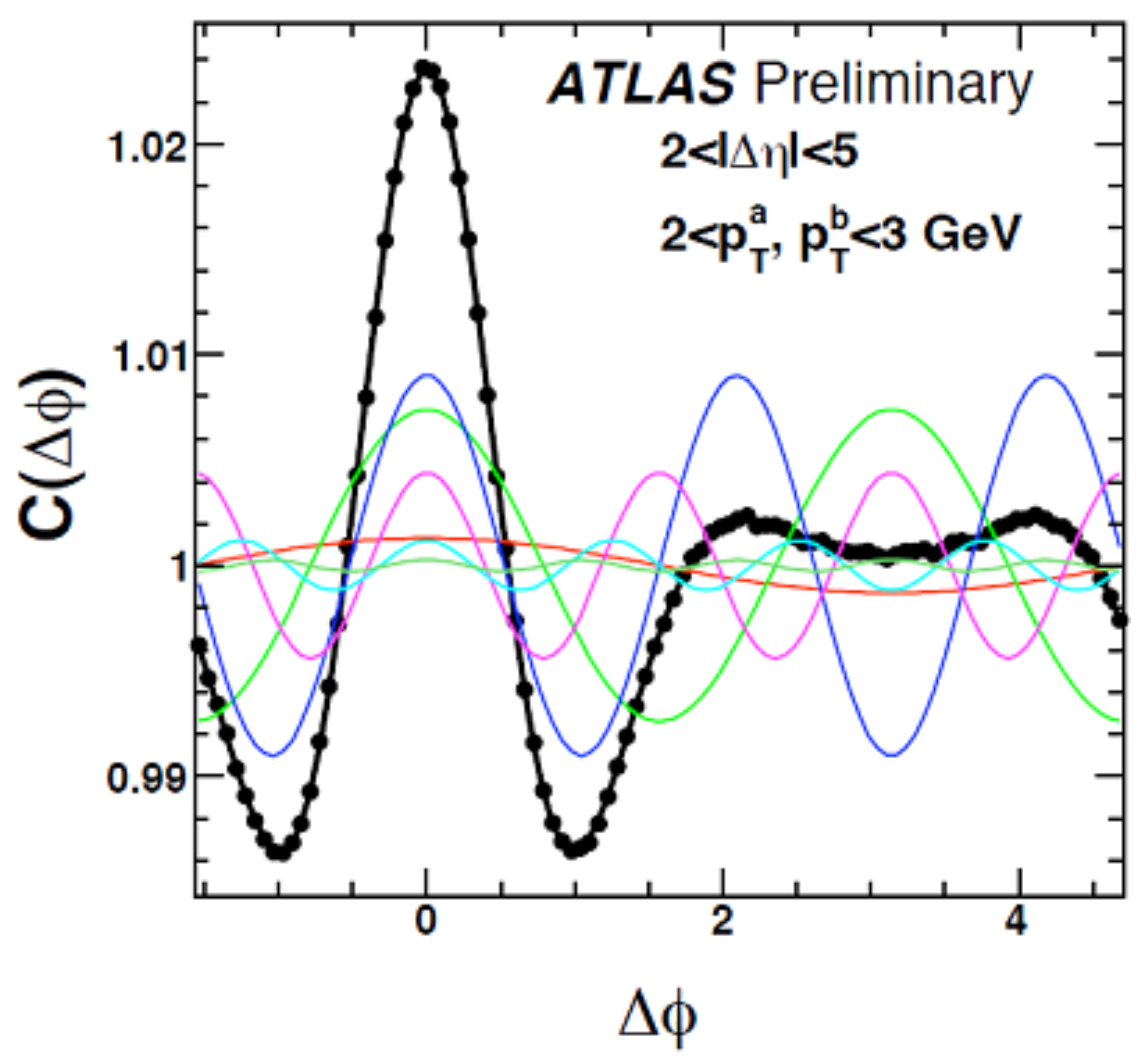}
\caption{Top: Two particle correlations (arbitrary units) as a function of $\Delta\phi$ obtained for an initial Gaussian perturbation of width 0.7 fm for the ideal case (left) and for the case with $\eta/s=0.16$ (right). Bottom: Two particle correlation (arbitrary units) as a function of $\Delta\phi$ obtained for an initial Gaussian perturbation of width 1 fm for the ideal case (left). Two particle correlation function measured by the ATLAS collaboration \cite{Collaboration:2011hf} for $0-1\%$ centrality (right).\label{2pcorr}} 
\end{figure}

We would now like to qualitatively compare our results to some experimental data. In order to do so we calculate the two-particle correlation function by multiplying two single particle distributions $dN/d\phi$ from the previous section and averaging over the azimuthal position of the initial Gaussian perturbation.  Since we are studying only perfectly central collisions this result depends only on the difference $\Delta\phi=\phi_1-\phi_2$.  Our results are plotted together with data from ATLAS \cite{Collaboration:2011hf} in Fig.~\ref{2pcorr} for comparison. We present two plots calculated for different viscosities and for an initial perturbation of width 1 fm (top of Fig.~\ref{2pcorr}), where it is possible to see that the main features of the two-particle correlation function resemble what is measured experimentally. There is one large central peak at $\Delta\phi=0$ and centred around $\Delta\phi=\pi$ there is a double hump, and since we did not include any rapidity dependence in our perturbation, these structures are elongated in rapidity. 
Although there is a similarity between our results and the experiments, we must mention that they do not exactly match; in our case the flat zone centred at $\pi$ is more extended than in the data. To understand this it is important to remember that Gubser's flow that we use is an idealization, were matter is assumed to be conformal and to satisfy the equation of state $\epsilon=3p$, which implies a constant speed of sound $c_s=1/\sqrt{3}$ throughout the whole evolution. This doesn't happen in real heavy ion collisions, where the speed of sound varies.  The final radius of the sound circle will depend on the speed of sound and on the freeze-out time, thus the distance between the two bumps in the two particle correlation is also affected by these parameters. Since we use an idealized situation our results do not perfectly match the experiment, but it allows us to do a qualitative study and check that the picture of an expanding circle from an initial localized perturbation yields a similar result to the ones found in experiment.\\

The top left plot in Fig.~\ref{2pcorr}, for an initial Gaussian perturbation of width 0.7 fm and zero viscosity, presents some extra structure in comparison to the experimental data; next to the main peak at $\Delta\phi=0$ two smaller peaks appear.   However, if viscosity is included (top right plot) or the width of the initial perturbation is increased (bottom left plot) these extra peaks disappear.  As we will see below, both the initial width of the perturbation and the viscosity of matter affect large harmonics, which die for large values of these parameters, so we can conclude that the extra peaks are due to the large harmonics and that to match the experimental results  the widths of the initial perturbations  and/or the viscosity of the medium must be large.\\

\begin{figure}[ht]
\centering
\includegraphics[width=0.33\textwidth]{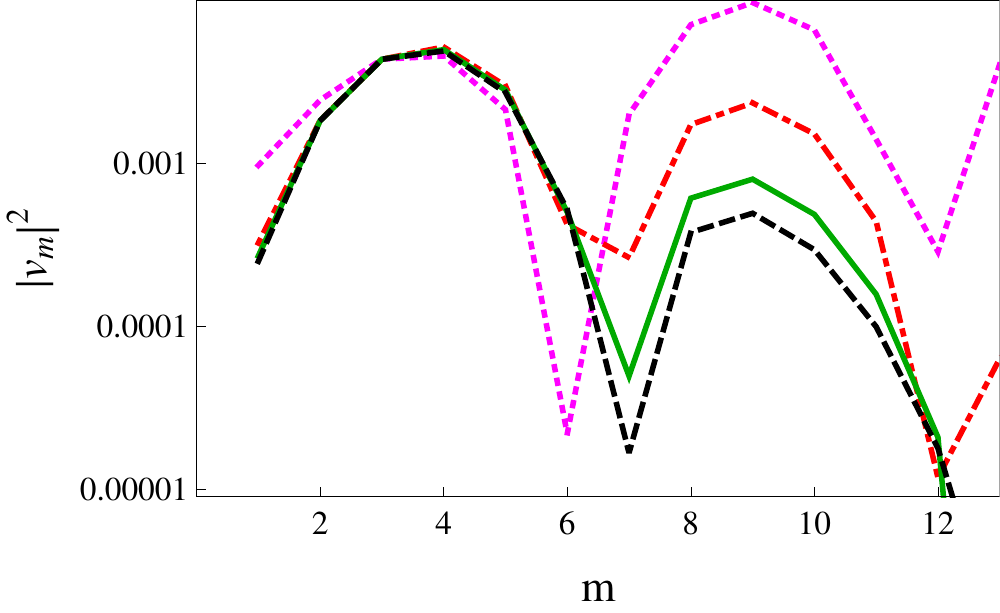}
\includegraphics[width=0.33\textwidth]{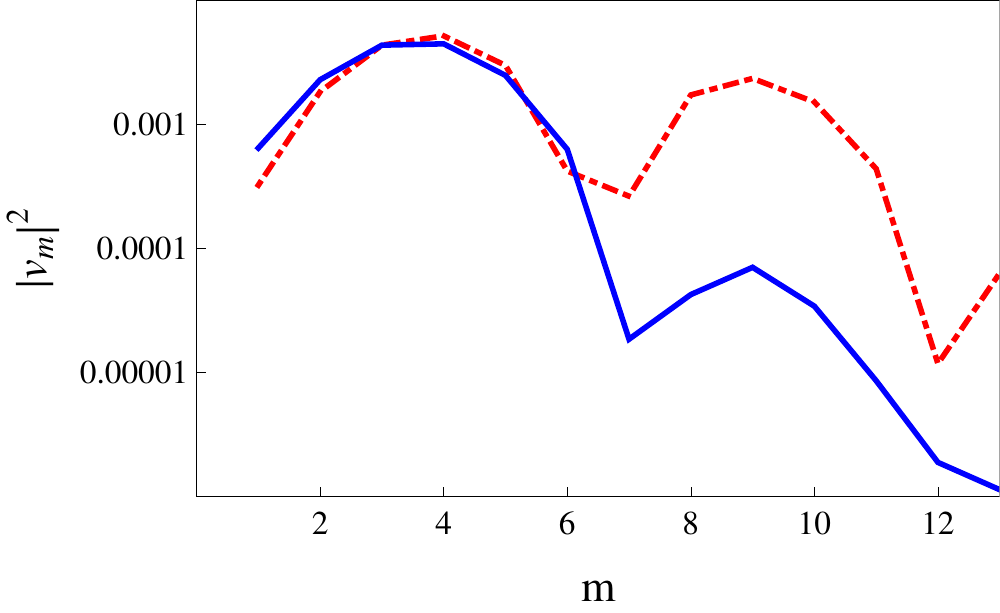}
\includegraphics[width=0.32\textwidth]{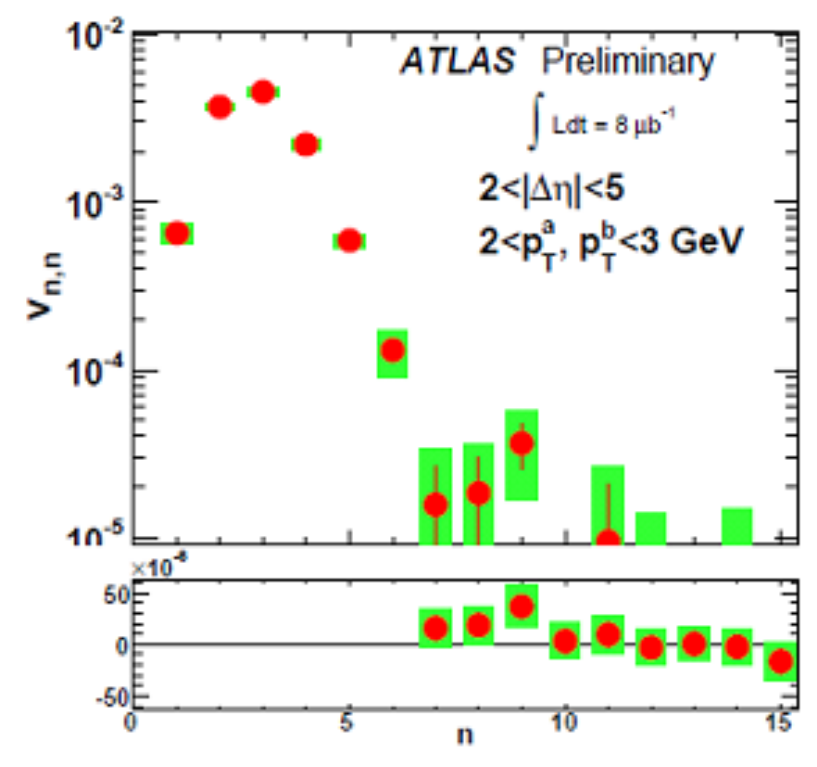}
\caption{Spectral plots $v_n^2$ vs $n$. Left: for viscosities $\eta/s=$0 (small dashed magenta),0.08(dot-dashed red),0.134 (solid green),0.16 (dashed black) and width of the initial Gaussian perturbation of 0.4 fm. Center: for widths of the initial Gaussian perturbation of 0.4 (dot-dashed red), 1 (solid blue) fm and viscosity $\eta/s=0.08$ .  Right:from the ATLAS collaboration \cite{atlas2}. Both the left and center plots have been normalized such that the value of v$_n^2$ for $n=3$ matches the data.\label{spec}} 
\end{figure}

The two particle correlation function can be expressed in terms of its Fourier expansion, given by
\begin{eqnarray}
\frac{dN}{d\Delta\phi} \propto 1+2\sum_n \textrm{v}_n^2 \cos{m\Delta\phi},
\end{eqnarray}
where the v$_n$ are the flow coefficients in the Fourier expansion of the single particle distribution. Of course, if we now plot v$_n$ vs $n$, we will be looking at the same information that is contained in the two particle correlation plots, but it is still interesting to look at the power spectrum and its features.  In Fig.~\ref{spec} it is possible to appreciate  that the power spectrum appears to have a first maximum, then starts to decay and then secondary maxima appear. This resembles what is seen in the Cosmic Microwave Background (CMB) radiation spectrum where after a first maximum some secondary maxima follow (see, for example\cite{Jarosik:2010iu}).
The plot on the left shows the dependence of the Fourier coefficients on viscosity, and it can be seen that, as we mentioned earlier, viscosity kills the higher harmonics.  The same effect is seen in the middle plot, this time we study the effect of the width of the initial Gaussian perturbation and we find that for larger initial widths the larger harmonics are suppressed in comparison to when smaller initial widths are used. This agrees with the study by Qin et al \cite{Qin:2010pf}, where it was shown how the larger widths of the initial Gaussian distributions suppressed the initial deformations $\epsilon_n$ with larger n.  Our results are to be compared to the plot from ATLAS \cite{atlas2}, which shows a large peak with a maximum at $n=3$, the curve then falls, but at $n=9$ a second maximum seems to appear.  This second maximum is much smaller than than the first one, implying that a large width of the initial Gaussian perturbation and/or a large viscosity are necessary.

\section{Summary} 
In this talk we presented the evolution of an initial state Gaussian deformation on top of the fireball, and the effects it has on measurable quantities such as the two particle correlations and the flow coefficients v$_n$. Using the formalism developed in \cite{Gubser:2010ui} by Gubser and Yarom, we were able to write the initial perturbation as a sum of the solutions to the hydrodynamic equations, and to see the evolution of the perturbation in time as it propagated from its center as an expanding circle, that by the freeze-out time had reached the edge of the expanding matter, and contributed two horns to the single particle distribution.\\

We looked at the two particle correlations and to the flow coefficients calculated for different widths of the initial Gaussian perturbation and for different viscosities. The variation of these two parameters proved to have similar results, the larger their values the smoother the particle distributions, and the more suppressed the values of the higher harmonics become. Comparing our results for the spectral plots with the data we find that for them to agree a large initial width of the perturbation and/or al large value of viscosity would be necessary, because the magnitude of the second peak is much smaller that that of the first.\\

It is necessary to mention that we worked with an idealized situation, where we looked only at the case with only one initial Gaussian deformation but in reality the initial state fluctuations are determined by the random positions of the nucleons in the nuclei and many such deformations are expected.  In order to take this into account it would be necessary to average over many initial deformations using probability  distributions to calculate their locations and amplitudes. Other assumptions that we used that do not hold in real heavy ion collisions are that we took the matter to be conformal throughout the process, and the initial perturbations to be small, in order to be able to use the analytical tools at hand. Still we find it remarkable, that even though we treated a very idealized situation in which we assumed that all harmonics add up coherently  to form the expanding circle from the perturbation, the results that we obtained agree qualitatively quite well with experimental data.


\bigskip 

\end{document}